\documentclass[aps,prr,twocolumn,superscriptaddress]{revtex4-2}

\usepackage{graphicx}
\usepackage{xcolor}
\usepackage{dcolumn}
\usepackage{bm}
\usepackage{amsmath}
\usepackage{amssymb}

\begin{document}

\title{Dynamic modes of active Potts models with factorizable numbers of states}

\author{Hiroshi Noguchi}
\email[]{noguchi@issp.u-tokyo.ac.jp}
\affiliation{Institute for Solid State Physics, University of Tokyo, Kashiwa, Chiba 277-8581, Japan}
\date{\today}

\begin{abstract}
We studied the long-term nonequilibrium dynamics of $q$-state Potts models with $q=4$, $5$, $6$, and $8$ using Monte Carlo simulations
on a two-dimensional square lattice.
When the contact energies between the nearest neighbors for the standard Potts models are used,
cyclic changes in the $q$ homogeneous phases and $q$-state coexisting wave mode appear at low and high flipping energies, respectively, for all values of $q$.
However, for a factorizable  $q$ value, dynamic modes with skipping states emerge,
depending on the contact energies.
For $q=6$, a spiral wave mode with three domain types (one state dominant or two states mixed)
and cyclic changes in three homogeneous phases are found.
Although three states can coexist spatially under thermal equilibrium,
the scaling exponents of the transitions to the wave modes are modified from the equilibrium values.
\end{abstract}

\maketitle

\section{Introduction}

Various types of pattern formation have been observed in nonequilibrium systems~\cite{nico77,hake04,mikh94,rabi00,kura84,aceb05,murr03}.
Classically, they have been called self-organizing systems~\cite{nico77,hake04}, dissipative structures~\cite{nico77}, 
and synergetics~\cite{hake04,mikh94}, 
and recently, nonreciprocal systems~\cite{you20,fruc21,rana24,guis24}, featuring different aspects.
In biological systems, spatiotemporal patterns have been observed from microscopic to macroscopic scales (cells and tissues to
animal populations)~\cite{mikh94,murr03,beta17,kond21,bail22,nogu24c}. 
Under far-from-equilibrium conditions, the thermal fluctuations occurring in macroscopic patterns are typically negligible
and the dynamics can be explained using deterministic continuum equations.
In contrast, the thermal fluctuations can be significant in microscopic patterns, such as intracellular waves.
However, the effects of the fluctuations and nucleation are not understood so far.

Lattice models, including Ising and Potts models, have been widely used to study the phase transitions occurring at thermal equilibrium~\cite{wu82,pott52,baxt73,bind80,bind81,ditz80,gres81,plas86,bena88,kohm81,frad84,okub15,nish16,iino19,musi21}.
The nonequilibrium dynamics have been investigated by imposing periodic external fields~\cite{akin17,sant21}
and spatial temperature inhomogeneity~\cite{mart20}.
Recently, the cyclic flipping energies of three or four states
have been introduced to these lattice models, and their pattern formations have been studied~\cite{nogu24a,nogu24b,nogu25,mana25,avni25,avni25a}. We have used the standard Potts models~\cite{wu82,pott52} for three~\cite{nogu24a,nogu24b} and four~\cite{nogu25} states in a two-dimensional (2D) square lattice
under both cyclic symmetry~\cite{nogu24a,nogu25} and asymmetry~\cite{nogu24b,nogu25};
 Hence, the transitions between dynamic modes, such as the cyclic changes of homogeneous phases and spiral waves, were revealed.
These dynamics were also obtained in an off-lattice model for undulating fluid membranes~\cite{nogu25a}.
These models exhibit stable spatiotemporal patterns in the long-term limit, unlike lattice Lotka--Volterra models,
which exhibit an absorbing transition to a uniform state~\cite{szol14,szab02,reic07,szcz13,kels15,dobr18,szab04,szab08,roma12,rulq14,baze19,zhon22,yang23,szol23}.
Manacorda and Fodor reported wave and global oscillation modes using a lattice model, which allows the occupancy of multiple particles at each site~\cite{mana25}.
Further, Avni et al. have added a nonreciprocal interaction to  the Ashkin-Teller model with a pairwise spin interaction (it is also called four-state vector-Potts and clock models~\cite{wu82})
and revealed changes in the scaling exponents of the second-order transition between disordered and ordered phases~\cite{avni25,avni25a}. 
However, this ordered phase maintains only for a short period ($t \propto$ system size $N$) starting from a uniform initial state.

With the aim of clarifying the effects of thermal fluctuations on pattern formation and nonequilibrium phase transitions, this study investigates several transitions between the dynamic modes in the long-term limit ($t\to \infty$), which are independent of the initial states.
We use general $q$-state Potts models with  $q=4$, $5$, $6$, and $8$ under cyclically symmetric conditions.
Various dynamic patterns are formed, particularly for a factorizable $q$ number.
Although phases comprising factorized numbers of states have been reported at thermal equilibrium~\cite{ditz80,gres81,plas86,bena88,taka20}, to the best of our knowledge,
spatiotemporal patterns with factorized symmetry have not been reported.
We reveal several types of  patterns with factorized symmetries and transitions between them.

The model and methods are described in Sec.~\ref{sec:model}.
Simulation results are presented and discussed in Sec.~\ref{sec:results}.
The results for $q=6$, $5$, $4$, and $8$ are described in Secs.~\ref{sec:6},
~\ref{sec:5},~\ref{sec:4}, and \ref{sec:8}, respectively.
The relationship with other models is discussed in Sec.~\ref{sec:other}.
Finally, a summary is presented in Sec.~\ref{sec:sum}.

\section{Active Potts Models}\label{sec:model}

A $q$-state Potts model of a 2D square lattice with side length $L$ is considered. The total number of sites is $N=L^2$, and
each site has a state $s\in [0,q-1]$.
The nearest neighboring sites ($i$ and $j$) have contact energies $J_{s_i,s_j}$:
\begin{equation}
\label{eq:hint}
H_{\mathrm{int}} = - \sum_{\langle ij\rangle} J_{s_is_j}.
\end{equation}
In standard Potts models, $J_{s_i,s_j}= J_0\delta_{s_i,s_j}$ is used.
In equilibrium systems, each state can additionally have a self-energy $\varepsilon_s$,
and the ratio of the forward and backward flip rates is $\exp(-\Delta H_{s_is'_i})$ for  flipping of a single site from $s$ to $s'$.
The thermal energy $k_{\mathrm{B}}T$ is normalized to unity,
$\Delta H_{s_is'_i} = \Delta H_{\mathrm{int}} - h_{s,s'}$, and the flipping energy $h_{s,s'} = \varepsilon_s - \varepsilon_s'$.
The cyclic sum of the flipping energy is $\sum_{k=0}^q h_{k,[k+1]} = 0$ at equilibrium, where $[k]$ indicates $k \bmod q$.
We extended this model to a nonequilibrium situation, in which $\sum_{k=0}^q h_{k,[k+1]} > 0$, but $h_{ss'}=-h_{s's}$ remains~\cite{nogu24a}.
Hence, the detailed balance can be locally satisfied for flips between $s$ and $s'$, but not globally for cycles ($s=0 \to 1\ ... \to (q-1) \to 0$). For $q=3$, this corresponds to the rock--paper--scissors relationship.
Note that this situation can be realized by reactions on a catalytic surface~\cite{ertl08,bar94,goro94,barr20,zein22} and molecular transport through a membrane~\cite{tabe03,miel20,holl21,nogu23,nogu25a}.
In the context of surface reactions, the state $s=0$ represents an unoccupied site, $s=1$ is a reactant-bound state, and $s=2$ to $q-2$ and $s=q-1$ are intermediate and final products, respectively.
The bulk reaction energy for the final product  is expressed as $\sum_{k=0}^q h_{k,[k+1]}$.
In the case of molecular transport, the molecules have $q-1$ intermediate states in the membrane.

\begin{figure}[tbh]
\includegraphics[]{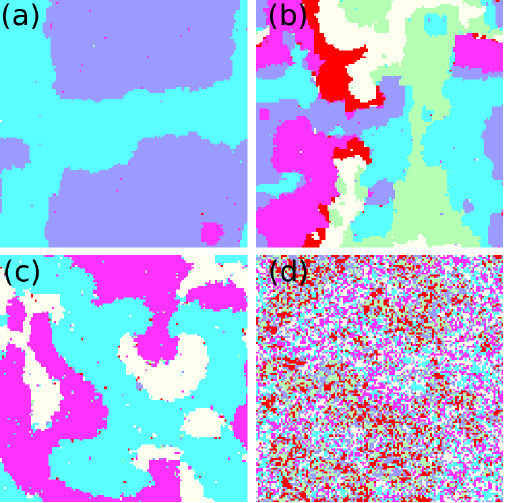}
\includegraphics[]{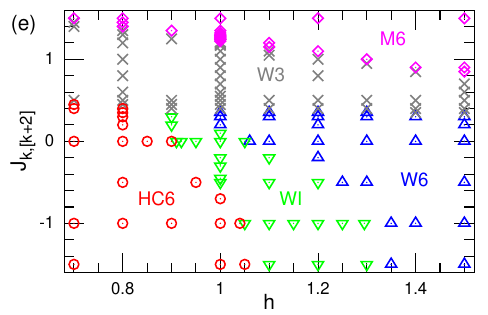}
\caption{
Six-state Potts model at $J_{k,[k+3]}=J_{k,[k+2]}$ and $L=128$ (Pot6a).
(a--d) Snapshots. 
The light yellow, light green, cyan, blue, magenta, and red sites (light to dark in grayscale) 
represent $s=0$, $1$, $2$, $3$, $4$, and $5$, respectively.
(a) Waves of $s=3$ and $s=4$ domains in the  intermediate wave mode (WI) at $h=1.2$ and $J_{k,[k+2]}=-1$.
(b) Waves of six states (W6) at  $h=1.5$ and $J_{k,[k+2]}=-1$.
(c) Waves of three states (W3) at $h=1$ and $J_{k,[k+2]}=0.5$.
(d) Mixing of six states (M6) at $h=1$ and $J_{k,[k+2]}=1.5$.
(e) Dynamic phase diagram.
The red circles, gray crosses, and magenta diamonds represent
HC6, W3, and M6, respectively.
The blue up-pointing triangles and  green down-pointing triangles
represent W6 and WI, respectively.
}
\label{fig:6d}
\end{figure}

\begin{figure}[tbh]
\includegraphics[]{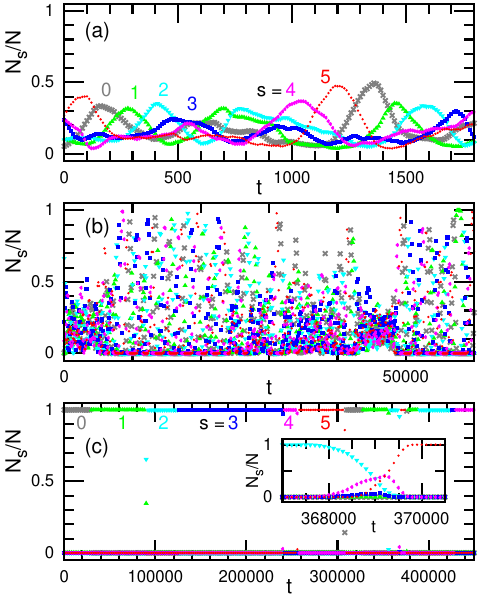}
\caption{
Time development of number densities of states in the six-state Potts model at $J_{k,[k+3]}=J_{k,[k+2]} = -1$ and $L=128$ (Pot6a).
(a) W6 at $h=1.5$. (b) WI at $h=1.2$. (c) HC6 at $h=0.8$.
(c, inset) Enlarged  density development at $t\sim 370~000$ showing  skipping of $s=3$ and $4$ phases.
}
\label{fig:td1}
\end{figure}

\begin{figure}[tbh]
\includegraphics[]{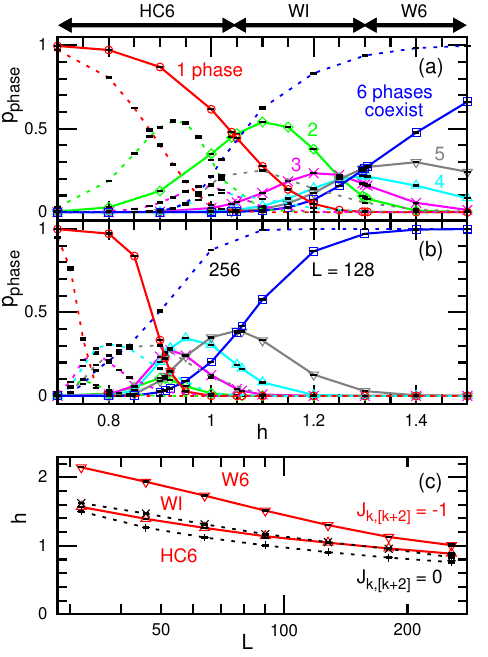}
\caption{
Dynamic modes of the six-state Potts model at $J_{k,[k+3]}=J_{k,[k+2]} \le 0$ (Pot6a).
(a--b) Probabilities $p_{\mathrm{phase}}$ of one-phase and multi-phase coexistence states as functions of $h$
at (a) $J_{k,[k+2]}=-1$ and (b) $J_{k,[k+2]}=0$.
The solid and dashed lines represent the data at $L=128$ and $256$, respectively.
The bidirectional arrows at the top represent the ranges of dynamic modes at $J_{k,[k+2]}=-1$.
(c) System-size dependence of phase boundaries.
The solid and dashed lines represent the data at $J_{k,[k+2]}=-1$ and $0$, respectively.
The statistical errors are shown as black bars. For most data points, they are smaller than the line thickness.
}
\label{fig:md01}
\end{figure}

\begin{figure}[tbh]
\includegraphics[]{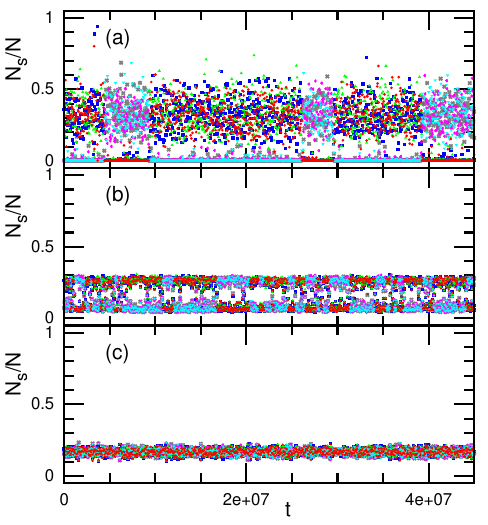}
\caption{
Time development of number densities of states in the six-state Potts model at $J_{k,[k+3]}=J_{k,[k+2]}$, $h=1$, and $L=128$ (Pot6a).
(a) W3 at $J_{k,[k+2]}=0.5$. (b) Transition point at $J_{k,[k+2]}=1.214$. (c) M6 at $J_{k,[k+2]}=1.5$.
}
\label{fig:tm1d}
\end{figure}

This study focuses on the dynamics occurring under the cyclic symmetry, such that
the flipping energies are constant as $h_{s,s'}=h$ and the contact energy $J_{s,s'}$ depends on the state distance $|s-s'|$ only.
A site is randomly selected and
the flip to the neighboring states ($s=k \to [k+1]$ or $[k-1]$) is performed using the Metropolis Monte Carlo (MC) algorithm.
This flip is performed $N$ times per MC step (time unit).
We have checked that the choice of the Metropolis and Glauber rates for the update process introduces no qualitative change in the dynamics at $q=3$~\cite{nogu24a}.

Through this study,
$J_{k,k}=2$ and $J_{k,[k+1]}=0$,
with $h$ and the other contact energies being varied:
 $J_{k,[k+2]}$ for $q=4$ and $5$,  $J_{k,[k+2]}$ and $J_{k,[k+3]}$ for $q=6$, and $J_{k,[k+2]}$, $J_{k,[k+3]}$, and $J_{k,[k+4]}$ for $q=8$ ($J_{k,k'}=J_{k',k}$).
For the $q$-state standard Potts models in 2D under thermal equilibrium,
the disorder-order transition is continuous and first-order for $q\le 4$ and $q>4$,
respectively,
and the transition point is given by $J_{0,\mathrm{c}}= \ln(1+ \sqrt{q})$~\cite{wu82,baxt73}.
Hence, the contact energy in this study, $J_{k,k}=2$, is sufficiently high for ordered phases in the standard Potts models to be obtained.
The statistical errors were calculated from three or more independent runs (ten runs were used to calculate the scaling exponents).

\section{Results and Discussion}\label{sec:results}
\subsection{Six-State Potts Model}\label{sec:6}

\subsubsection{Modes for $J_{k,[k+3]}=J_{k,[k+2]}$}\label{sec:6d}

\begin{figure}[h!]
\includegraphics[]{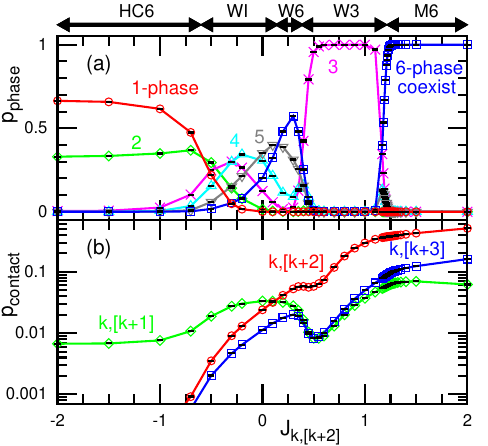}
\caption{
Dependence on $J_{k,[k+2]}$ in the six-state Potts model at  $J_{k,[k+3]}=J_{k,[k+2]}$, $h=1$, and $L=128$ (Pot6a).
(a) Probabilities $p_{\mathrm{phase}}$ of one-phase and multi-phase coexistence states.
The bidirectional arrows at the top represent the ranges of dynamic modes.
(b) Contact probabilities $p_{\mathrm{contact}}$ for $k$--$[k+1]$, $k$--$[k+2]$, and $k$--$[k+3]$ pairs.
}
\label{fig:m1d}
\end{figure}

\begin{figure*}[tbh]
\includegraphics[]{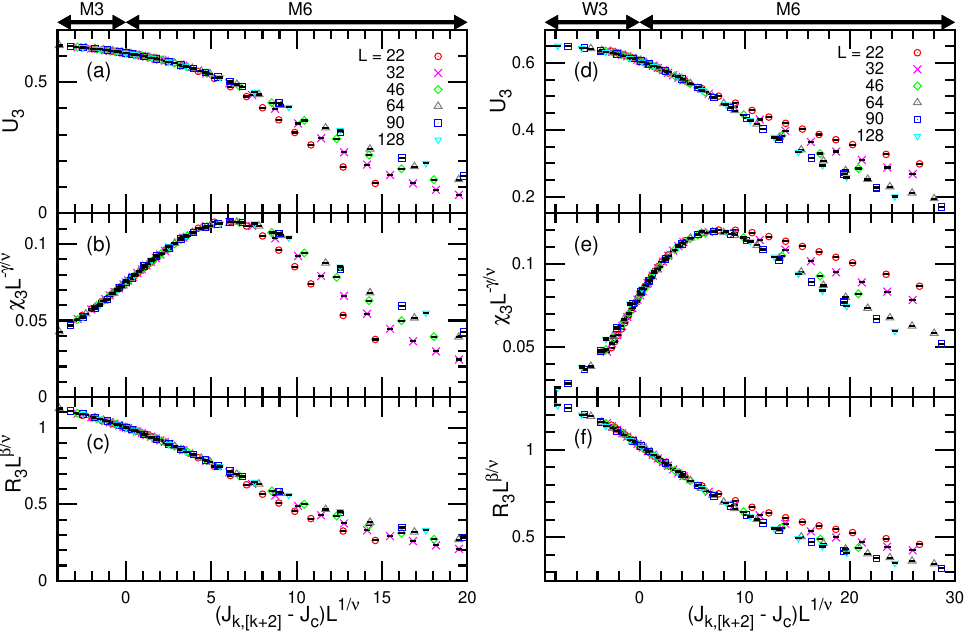}
\caption{
Transition between W3 (M3) and M6 modes at $J_{k,[k+3]}=J_{k,[k+2]}$
for (a--c) $h=0$ and (d--f) $h=1$ (Pot6a).
Binder cumulant $U_3$, susceptibility $\chi_3$, and order parameter $R_3$ for three-fold rotational symmetry
are shown in (a,d), (b,e), and (c,f), respectively.
The transition points are  $J_{\mathrm{c}}=1.725$  and $1.214$ for $h=0$ and $1$, respectively.
}
\label{fig:m01dc}
\end{figure*}

For six-state Potts model ($q=6$),
we have a 3D parameter space ($h, J_{k,[k+2]}, J_{k,[k+3]}$), while keeping $J_{k,k}=2$ and $J_{k,[k+1]}=0$.
We simulated the dynamics using parameters in three 2D slices ($J_{k,[k+3]}=J_{k,[k+2]}$, $J_{k,[k+3]}=0$, and $J_{k,[k+2]}=0$) to widely survey the parameter space.
First, we describe the dynamics for $J_{k,[k+3]}=J_{k,[k+2]}$ (Figs.~\ref{fig:6d}--\ref{fig:m01dc}).
We refer to this condition as Pot6a.
Five modes are obtained as shown in Fig.~\ref{fig:6d}.

For $J_{k,[k+2]}\leq 0$, negative contact energies are obtained only between neighbor pairs of the same states (i.e., repulsion between different states).
Hence, a single state of $s\in [0,5]$ dominantly spans the entire lattice at $h=0$, and these six homogeneous phases are equivalent.
At low but nonzero $h$, these six homogeneous phases change cyclically as $s=0\to 1 \to 2\to 3 \to 4 \to 5 \to 0$ 
(see Fig.~\ref{fig:td1}(c)).
This is a homogeneous cycling mode, as previously observed in the three- and four-state active standard Potts models~\cite{nogu24a,nogu25}. 
In this paper, this mode is abbreviated as HC6 (that is, HC$n$ denotes the cycling of $n$ states).
Note that cyclic changes occur through nucleation and growth, and the nucleation period decreases exponentially with increasing $h$~\cite{nogu24a}.
When a nucleus of the next state ($s=[k+1]$) appears in the growth of an $s=k$ domain,
the  $s=k$ dominant phase is skipped~\cite{nogu25}.
Skipping of the $s=3$ and $4$ phases is shown in
Fig.~\ref{fig:td1}(c, inset).
As $h$ increases, the subsequent nuclei form more frequently during domain growth (Fig.~\ref{fig:6d}(a));
 subsequently, spatial coexistence of multiple states becomes dominant.
At high $h$, all six states spatially coexist, and the domain boundary between neighboring states $s=k$ and $[k+1]$  moves ballistically
in the direction  from the $s=[k+1]$ to $s=k$ domains (see Fig.~\ref{fig:td1}(a) and Movie S1).
Herein, this mode is abbreviated as W6 (where W$n$ denotes the waves of $n$ states).
With increasing repulsion $|J_{k,[k+2]}|$, the boundary length between the $s=k$ and $[k+2]$ or $[k+3]$ domains decreases,
such that the waves more frequently form a spiral shape (see Movie S1).
At intermediate values of $h$, intermediate numbers of states ($2$--$5$) spatially coexist for most of the period (see the bottom region in Fig.~\ref{fig:td1}(b). A few states remain at $N_s/N\simeq 0$ for most of the period). 
In this study, this intermediate wave mode is abbreviated as WI.
This behavior differs from those in three- and four-state active standard Potts models~\cite{nogu24a,nogu25},
in which HC$q$ and W$q$ modes temporally coexist in the intermediate conditions.
For $q=6$, HC6 and W6 modes  temporally appear in the WI mode but only for short periods, such that the lattice is primarily occupied by intermediate numbers of states.

\begin{figure}[tbh]
\includegraphics[]{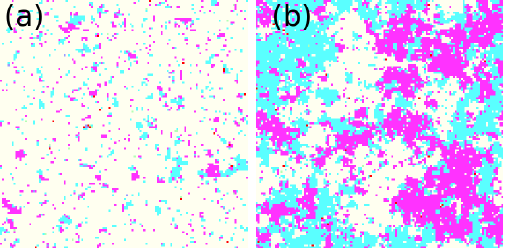}
\includegraphics[]{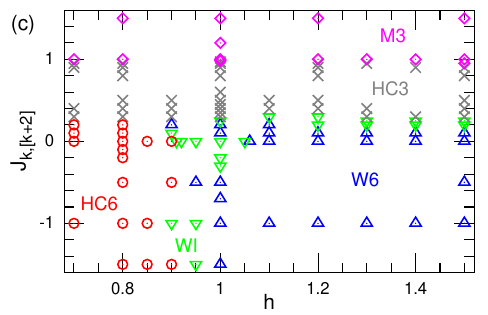}
\caption{
Six-state Potts model with $J_{k,[k+3]}=0$ and $L=128$ (Pot6b).
(a) Snapshot of homogeneous cycling of three states (HC3) at  $h=1$ and $J_{k,[k+2]}=0.95$.
(b) Snapshot of mixing of three states (M3) at $h=1$ and $J_{k,[k+2]}=1$.
(c) Dynamic phase diagram.
The red circles, gray crosses, magenta diamonds, blue up-pointing triangles, 
and  green down-pointing triangles represent
HC6, HC3, M3, W6, and WI, respectively.
}
\label{fig:6y}
\end{figure}

\begin{figure}[tbh]
\includegraphics[]{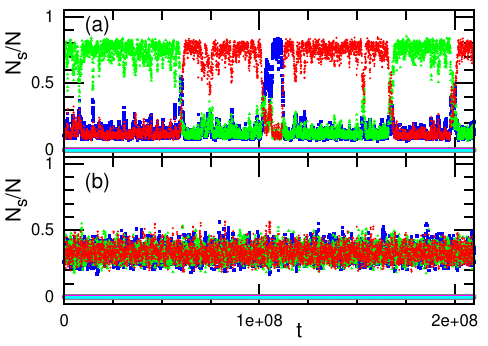}
\caption{
Time development of number densities of states in the six-state Potts model at $J_{k,[k+3]}=0$ and $h=1$ (Pot6b).
(a) HC3 at $J_{k,[k+2]}=0.976$. (b) M3 at $J_{k,[k+2]}=1$.
}
\label{fig:tm1y}
\end{figure}

To distinguish between these three modes, we calculate the time fractions $p_{\mathrm{phase}}$ of single phases and multi-phase coexistence.
It is considered that the lattice is covered by one phase at $N_s/N>0.95$ for $s \in [0,q-1]$,
and that $n$ phases spatially coexist when $n$ states satisfy $N_s/N>0.05$.
The fractions of the one-phase dominance and six-phase coexistence decreases and increases with increasing $h$, respectively
(see Fig.~\ref{fig:md01}(a) and (b)).
For the HC and W6 modes,
the largest fractions are obtained for the one-phase dominance and six-phase coexistence, respectively.
For the WI mode,
the largest fraction is obtained for
 one of the $n$-phase coexistences, where $n=2$--$5$ (see the bidirectional arrows at the top of Fig.~\ref{fig:md01}(a)).
As the system size increases, the transition points of $h$ between HC6 and WI and between WI and W6 decrease (see Fig.~\ref{fig:md01}),
similar to those between HC$q$ and W$q$ in the three- and four-state active standard Potts models~\cite{nogu24a,nogu25}.

For $J_{k,[k+2]}> 0$, two additional modes emerge:
wave modes of three states (W3) and a mixed phase of six states (M6), as shown in Figs.~\ref{fig:6d}(c--e) and \ref{fig:tm1d}.
As $J_{k,[k+2]}$ increases, the length of the boundary between the $s=k$ and $[k+2]$ domains increases owing to a reduction in the interfacial tension
(see Fig.~\ref{fig:m1d}).
When $J_{k,[k+2]} \simeq J_{k,k}=2$, the energy losses due to the contact between the different states (excluding the neighboring states $s=k$ and $[k+1]$) become negligibly small; hence, in the M6 mode, the six states are mixed to gain entropy.
Note that this mode corresponds to the disordered phase at thermal equilibrium.
For the intermediate $J_{k,[k+2]}$, three odd- or even-numbered states ($s=1,3,5$ or $s=0,2,4$) form domains exhibiting spiral waves, as observed in the three-state Potts model~\cite{nogu24a,nogu24b} (see Fig.~\ref{fig:6d}(c) and Movie S2).
The other three states are present to small degrees at the domain boundaries.
The boundary of the $s=k$ and $[k+2]$ domains  moves via two-step flips ($s=k\to [k+1]\to [k+2]$),
in which the clusters of $s=[k+1]$ states are swallowed up by the $[k+2]$ domain before growing.
Waves of odd- and even-numbered states stochastically switch (see Fig.~\ref{fig:tm1d}(a) and (b)).
Note that this mode can be elucidated via symmetry factorization~\cite{taka20}.
Static factorized phases for $q=2^n$ have been reported for the standard Ashkin-Teller model ($n=2$)~\cite{ditz80,plas86,bena88} and extended $n$-color models~\cite{gres81}.
Here, the factorized (three-fold) symmetry remains within a $6=2\times 3$ symmetric system.
For the wave propagation modes, the three-fold symmetry helps maintain their spiral centers~\cite{nogu25}.

\begin{figure}[tbh]
\includegraphics[]{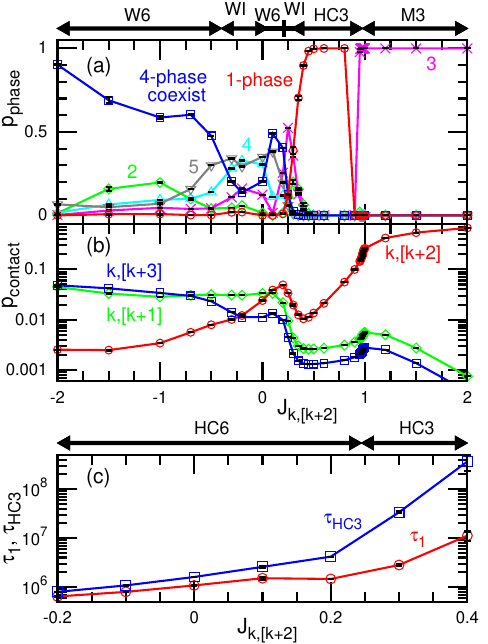}
\caption{
Dependence on $J_{k,[k+2]}$ in the six-state Potts model at $J_{k,[k+3]}=0$ and $L=128$
for (a--b) $h=1$ and (c)  $h=0.8$ Pot6b).
(a) Probabilities $p_{\mathrm{phase}}$ of one-phase and multi-phase coexistence states.
(b) Contact probabilities $p_{\mathrm{contact}}$ for $k$--$[k+1]$, $k$--$[k+2]$, and $k$--$[k+3]$ pairs.
(c) Lifetimes of one-phase state ($\tau_{1}$) and HC3 mode ($\tau_{\mathrm{HC3}}$).
The bidirectional arrows at the top represent the ranges of dynamic modes.
}
\label{fig:m1y}
\end{figure}

The transition between the W3 and M6 modes is continuous.
In the middle of the W3 range (Fig.~\ref{fig:tm1d}(a)), the mean densities of the three dominant states are approximately $1/3$, and those of the other three states are approximately $0$.
With increasing $J_{k,[k+2]}$, the mean densities of the dominant and minor states decrease and increase, respectively.
In the M6 mode, these mean densities converge to $1/6$ (see Fig.~\ref{fig:tm1d}(b) and (c)).
The transition between three and six states (M3 and M6) also occurs at thermal equilibrium ($h=0$),
and the transition point $J_{k,[k+2]}=J_{\mathrm{c}}$ is weakly dependent on $h$ (see Fig.~\ref{fig:6d}(e)).
The transition is characterized by the order parameter $R_3$ for the three-fold symmetry.
$R_n$ is defined as
\begin{eqnarray}
  R_n &=& \langle s_n \rangle, \\
   s_n  &=&\frac{1}{N}\bigg|\sum_j^N \exp\Big(\frac{2n\pi{\mathrm{i}}s_j}{q}\Big)\bigg|,
\end{eqnarray}
where $n$ is a positive integer and $\langle ... \rangle$ is the long-term time average.
When all the sites are occupied by one state, $R_n=1$ for any $n$.
When the sites are equally occupied by the three odd-numbered states,
$R_1=0$ and $R_3=1$.
When the sites are equally occupied by all six states,
$R_1=R_3 =0$.
The susceptibility $\chi_n$ and Binder cumulant $U_n$
are defined, respectively, as~\cite{bind81,avni25a}
\begin{eqnarray}
  \chi_n &=& N(\langle {s_n}^2 \rangle - \langle s_n \rangle^2), \\
   U_n  &=& 1 - \frac{\langle {s_n}^4 \rangle}{3\langle {s_n}^2 \rangle^2}.
\end{eqnarray}
For second-order transitions,
the scaling relations of the order parameter (here, $R_3$) are obtained in the vicinity of the transition point $J_{\mathrm{c}}$:
$R_3L^{\beta/\nu}= \tilde{R}_3[L^{1/\nu}(J_{k,[k+2]}-J_{\mathrm{c}}]$,
$\chi_3L^{-\gamma/\nu} = \tilde{\chi}_3[L^{1/\nu}(J_{k,[k+2]}-J_{\mathrm{c}})]$, and
$U_3 = \tilde{U}_3[L^{1/\nu}(J_{k,[k+2]}-J_{\mathrm{c}})]$,
where $\tilde{R}_3(x)$, $\tilde{\chi}_3(x)$, and $\tilde{U}_3(x)$
are the scaling functions and $\nu$, $\beta$, and $\gamma$ are the critical exponents.
Figure \ref{fig:m01dc} shows the normalized $U_3$, $\chi_3$, and $R_3$ as functions of $(J_{k,[k+2]}-J_{\mathrm{c}})L^{1/\nu}$ for $h=0$ and $1$.
We fitted each curve with a quartic function and calculated the scaling exponents by least-squares fits to these functions for $L \le 90$ in the vicinity of the transition point.

We obtained the exponents shown in Table~\ref{tab:expo} with
$J_{\mathrm{c}}= 1.725 \pm 0.002$ and $1.214 \pm 0.001$  for $h=0$ and $1$, respectively.
The exponent $\nu$ is close to unity and is smaller for $h=1$ than for $h=0$ (equilibrium).
Therefore, the exponents for the transition from M6 to the wave mode (W3) are modified from the equilibrium values between the mixed phases (M6 and M3).
Note that, in Fig.~\ref{fig:m01dc}, for $(J_{k,[k+2]}-J_{\mathrm{c}})L^{1/\nu} \gtrsim 10$, the curves do not converge well for either $h=0$ or $1$.
This deviation may be caused by the crossover effects~\cite{ditz80} from other transitions (e.g., the W6--W3 transition).

\begin{table}
\caption{\label{tab:expo} 
Scaling exponents for the transitions between W3(M3) and M6 (Pot6a) and between HC3(H) and M3 (Pot6b)
for $h=0$ and $1$.}
\begin{center}
\begin{tabular}{llll}
\hline
               & \ $1/\nu$  & \ $\gamma/\nu$  & \ $\beta/\nu$ \\
 \hline
Pot6a &&&\\
$h=0$\hspace{0.4cm}  &  $0.95 \pm 0.02$\hspace{0.4cm}  & $1.742 \pm 0.002$\hspace{0.4cm}  &  $0.1233 \pm 0.0004$  \\
$h=1$ &  $1.12 \pm 0.02$ & $1.732 \pm 0.007$ &  $0.132 \pm 0.002$ \\
Pot6b &&&\\
$h=0$ &  $1.24 \pm 0.05$ & $1.75 \pm 0.02$ &  $0.137 \pm 0.002$ \\
$h=1$ & $1.19 \pm 0.04$ & $1.761 \pm 0.008$ & $0.126 \pm 0.005$ \\
\hline
\end{tabular}
\end{center}
\end{table}

\begin{figure*}[tbh]
\includegraphics[]{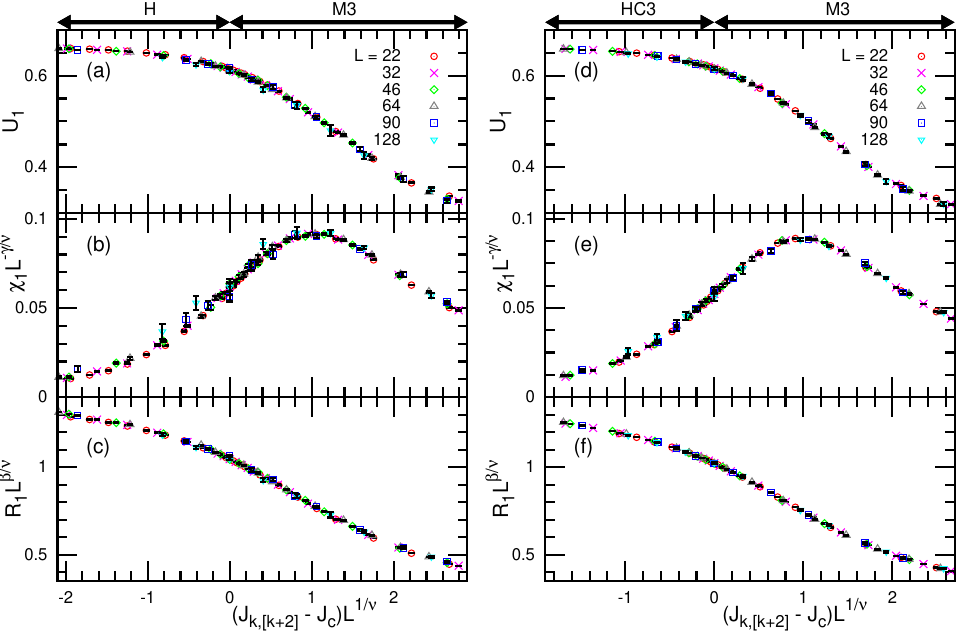}
\caption{
Transition between HC3 (H) and M3 modes at $J_{k,[k+3]}=0$
for (a--c) $h=0$ and (d--f) $h=1$ (Pot6b).
Binder cumulant $U_1$, susceptibility $\chi_1$, and order parameter $R_1$ on the single-phase occupancy
are shown in (a,d), (b,e), and (c,f), respectively.
The transition points are  $J_{\mathrm{c}}=0.992$  and $0.977$ for $h=0$ and $1$, respectively.
}
\label{fig:m01yc}
\end{figure*}

\subsubsection{Modes for $J_{k,[k+3]}=0$}\label{sec:6y}

Next, we investigate the $J_{k,[k+2]}$ dependence while keeping $J_{k,[k+3]}=0$ (Figs.~\ref{fig:6y}--\ref{fig:m01yc}).
We refer to this condition as Pot6b.
Although the dynamics for $J_{k,[k+2]} \le 0$ are similar to those for $J_{k,[k+3]}=J_{k,[k+2]}$,
two modes (HC3 and M3) are found for $J_{k,[k+2]} > 0$, instead of the W3 and M6 modes.

At $J_{k,[k+2]} = J_{k,k}=2$, three odd-numbered states ($s=1,3,5$)
can be randomly mixed without loss of contact energy in the absence of the even-numbered states;
similarly for the even-numbered states.
By contrast, mixing of the six states is inhibited by the repulsion between the diagonal pairs ($s=k$ and $[k+3]$) because $J_{k,[k+3]}=0$.
Thus, the mixed phases of the three states (M3) are stable for $J_{k,[k+2]}\simeq 2$,
for which the boundaries of the clusters of each state are not smooth and move diffusively rather than ballistically (see Figs.~\ref{fig:6y}(b), (c), and \ref{fig:tm1y}(b), and Movie S3).
No switching between the odd- and even-numbered states was observed during the available simulation periods $\sim 10^9$. This phase also exists at $h=0$.

For the intermediate values of $J_{k,[k+2]}$,
homogeneous cycling of three phases ($s=1 \to 3 \to 5$ or $s=0 \to 2 \to 4$)
occurs (HC3). 
The $s=k$ dominant phase involves small clusters of $s=[k+2]$ and $[k+4]$ states (see Figs.~\ref{fig:6y}(a) and \ref{fig:tm1y}(a)).
Hence, the $s=[k+1]$ dominant phase is skipped, because
the $s=[k+2]$ phase grows rapidly from these clusters
during the nucleation and growth of the $s=[k+1]$ phase.

The transition between the HC3 and M3 modes is continuous.
As $J_{k,[k+2]}$ increases in the HC3 range,
the densities of the dominant phase and the other two phases of the three major states
decreases and increase, respectively (see Fig.~\ref{fig:tm1y}).
Additionally, the contact probability $p_{\mathrm{contact}}$ between these three states increases (see Fig.~\ref{fig:m1y}(b)).
Hence, the three densities converge as the system approaches the M3 mode.
This change can be characterized by the order parameter $R_1$.
At $h=0$ (equilibrium), it becomes the transition between the M3 and one-state dominant phases (called H).
We calculated the scaling exponents using the method described in Sec.~\ref{sec:6d}.
Figure \ref{fig:m01yc} shows the normalized $U_1$, $\chi_1$, and $R_1$ as functions of $(J_{k,[k+2]}-J_{\mathrm{c}})L^{1/\nu}$ for $h=0$ and $1$.
We obtained the exponents as shown in Table~\ref{tab:expo} with
$J_{\mathrm{c}}= 0.992 \pm 0.001$ and $0.977 \pm 0.001$ for $h=0$ and $1$, respectively.
These values are close to the $\nu=5/6$, $\gamma=13/9$, and $\beta = 1/9$ obtained for the three-state standard Potts model~\cite{wu82}.
The differences in the exponents for $h=0$ and $1$ are in the ranges of statistical errors,
thus, the scaling exponents are not or little modified from the equilibrium values.
This outcome may reflect the static nature of the two modes.
That is, in most periods of the HC3 mode, a single-state dominant phase exists, similar to that obtained at equilibrium for $h=0$. 
The M3 mode is almost identical to the mixed phase at equilibrium.
To obtain a distinct change in the scaling exponents, dynamic modes, such as wave propagation in W3, may be required.

Under low $h$, the HC3 mode changes to the HC6 mode with decreasing $J_{k,[k+2]}$.
This change is continuous and can be characterized by the ratio of the lifetimes (one vs. three phases)
and the distribution of the state densities.
The state-density probability distribution $P(N_s/N)$ has three peaks in the HC3 mode (see Fig.~\ref{fig:tm1y}(a)); 
in the HC6 mode, the two smaller peaks merge into a single peak.
The lifetime $\tau_{1}$ of one phase is the average of the periods, in which one of the states ($s\in [0,q-1]$) maintains the condition of $N_s/N>0.95$.
The HC3 mode of the odd numbers is taken to be initiated when $N_{\mathrm{odd}}/N$ exceeds $0.95$
and to end when $N_{\mathrm{even}}/N$ exceeds $0.95$, where $N_{\mathrm{odd}}=N_1+N_3+N_5$ and $N_{\mathrm{even}}=N_0+N_2+N_4$.
Similarly, for that of the even numbers.
The lifetime $\tau_{\mathrm{HC3}}$ of the HC3 mode is the average of these two periods.
With increasing $J_{k,[k+2]}$, 
 $\tau_{1}$ gradually increases. However, $\tau_{\mathrm{HC3}}$ rapidly increases for $J_{k,[k+2]} \gtrsim 0.3$ at $h=0.8$ (see Fig.~\ref{fig:m1y}(c)).
We consider that the HC3  mode is determined by  $\tau_{\mathrm{HC3}}/\tau_{1}>10$,
since $\tau_{\mathrm{HC3}}$ is longer than $\tau_{1}$ in the HC3 mode.

\begin{figure}[tbh]
\includegraphics[]{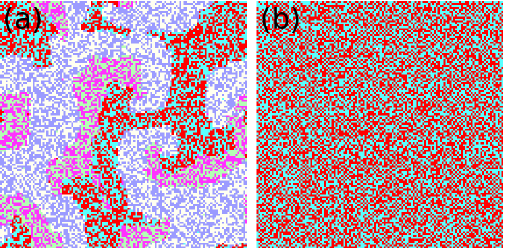}
\includegraphics[]{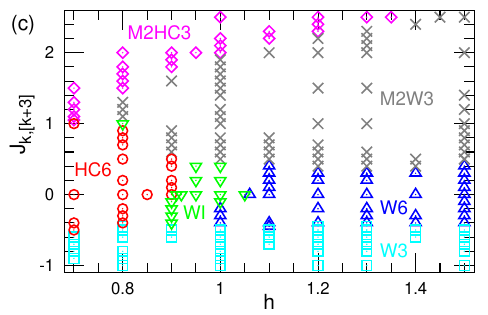}
\caption{
Six-state Potts model at $J_{k,[k+2]}=0$ and $L=128$ (Pot6c).
(a) Snapshot of spiral waves of three types of two-state mixed domains (M2W3) 
at  $h=1.3$ and $J_{k,[k+3]}=2$.
(b) Snapshot of homogeneous cycling of mixed phases (M2HC3) at $h=1.3$ and $J_{k,[k+3]}=2.5$.
Two diagonal states ($s=2$ and $5$) are mixed.
(c) Dynamic phase diagram.
The red circles, gray crosses, magenta diamonds, blue up-pointing triangles, 
green down-pointing triangles, and light blue squares represent
HC6, M2W3, M2HC3, W6, WI, and W3, respectively.
}
\label{fig:6t}
\end{figure}

\begin{figure}[tbh]
\includegraphics[]{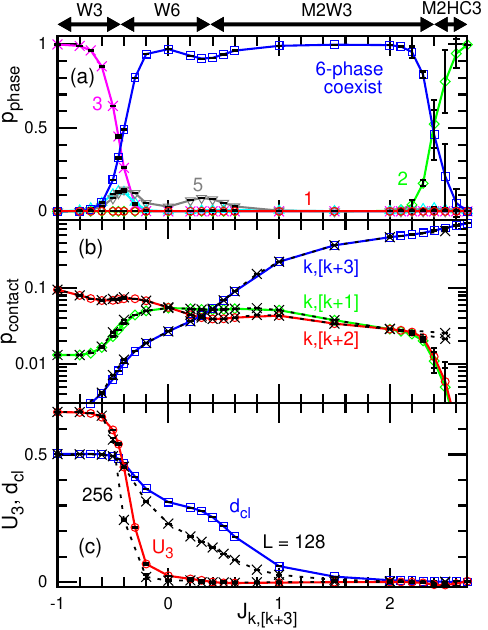}
\caption{
Dependence on $J_{k,[k+3]}$ in the six-state Potts model at  $J_{k,[k+2]}=0$ and $h=1.3$ (Pot6c).
(a) Probabilities $p_{\mathrm{phase}}$ of one-phase and multi-phase coexistence states.
The bidirectional arrows at the top represent the ranges of dynamic modes.
(b) Contact probabilities $p_{\mathrm{contact}}$ for $k$--$[k+1]$, $k$--$[k+2]$, and $k$--$[k+3]$ pairs.
(c) Binder cumulant $U_3$ for three-fold symmetry and cluster size ratio $d_{\mathrm{cl}}$.
The solid lines represent the data at $L=128$.
In (b) and (c),
the dashed lines with crosses represent the data at $L=256$.
}
\label{fig:m13t}
\end{figure}

\begin{figure}[tbh]
\includegraphics[]{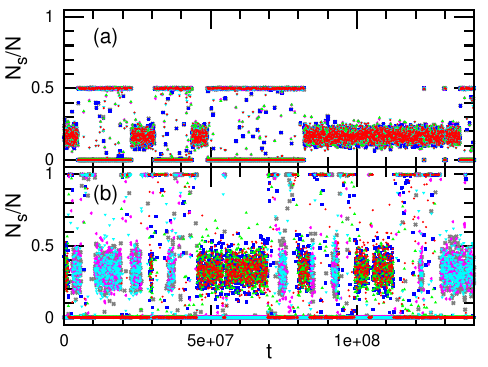}
\caption{
Time development of number densities of states in the six-state Potts model at $J_{k,[k+2]}=0$ and $L=128$ (Pot6c).
(a) Temporal coexistence of M2W3 and M2HC3 at $h=1.3$ and $J_{k,[k+3]}=2.4$. 
(b) Temporal coexistence of W3 and HC6  at $h=0.7$ and $J_{k,[k+3]}=-0.6$.
}
\label{fig:tmt}
\end{figure}

\subsubsection{Modes for $J_{k,[k+2]}=0$}\label{sec:6t}

Next, we investigate the $J_{k,[k+3]}$ dependence while keeping $J_{k,[k+2]}=0$ (Figs.~\ref{fig:6t}--\ref{fig:tmt}).
We refer to this condition as Pot6c.
For $J_{k,[k+3]} < 0$, the W3 mode emerges in addition to the HC6, WI, and W6 modes,
since the contact between the $k$ and $[k+2]$ states becomes more stable than that between the $k$ and $[k+3]$ states
 (see Figs.~\ref{fig:6t}(c) and \ref{fig:m13t}).
The transition between the W3 and HC6 modes occurs via the temporal coexistence of the two modes,
as shown in Fig.~\ref{fig:tmt}(b), similar to that occurring for the transition from HC3 to W3
in the three-state Potts model~\cite{nogu24a,nogu24b}.
The ratio of the W3 modes monotonously increases with decreasing $J_{k,[k+3]}$, 
and we determine the modes from the ratio of the phase periods.

For $J_{k,[k+3]} > 0$ two additional modes (M2W3 and M2HC3) are found.
At $J_{k,[k+3]} = J_{k,k}=2$, 
random mixing of two diagonal states ($s=k$ and $[k+3]$)
can occur without loss of contact energy in the absence of the other four states.
Hence, three types of mixed phases comprising two diagonal states exist with equal stability under $h=0$.
For $J_{k,[k+3]}> J_{k,k}$, diagonal-state contacts are preferred over those of the same states, as in the case for antiferromagnetic interactions in spin systems.
Thus, for $J_{k,[k+3]} \simeq J_{k,k}$, 
the homogeneous cycling and wave modes emerge for these three phases at low and high $h$, respectively, like in the three-state Potts model.
We call them M2HC3 and M2W3 (i.e., M$m$HC$n$ and  M$m$W$n$ for $n$ phases of $m$ mixed phases).
The transition between the M2W3 and M2HC3 modes occurs via the temporal coexistence of the two modes (see Fig.~\ref{fig:tmt}(a)).
 
In the M2W3 mode, the three types of mixed domains form spiral waves (see Fig.~\ref{fig:6t}(a) and Movie S4).
To characterize this mode, we calculated the $p_{\mathrm{contact}}$ values and cluster size ratio $d_{\mathrm{cl}}$.
Clusters of single or two diagonal states are considered.
Note that, in the latter case,
when neighboring sites have the same state or diagonal state ($s_i=k$ and $s_j=k$ or $[k+3]$), they belong to the same cluster.
With increasing $J_{k,[k+3]}$, the ratio $d_{\mathrm{cl}}=n_{\mathrm{c1}}/n_{\mathrm{c2}}$ of the mean sizes for these two clusters ($n_{\mathrm{c1}}$ and $n_{\mathrm{c2}}$ for the single and two diagonal states, respectively)
decreases and $p_{\mathrm{contact}}$ of the diagonal pairs ($s=k$ and $[k+3]$) increases  (see Fig.~\ref{fig:m13t}(b) and (c)).
We consider that the W6 mode changes to the M2W3 mode when the diagonal-pair $p_{\mathrm{contact}}$ exceeds the other probabilities.

\begin{figure}[h!]
\includegraphics[]{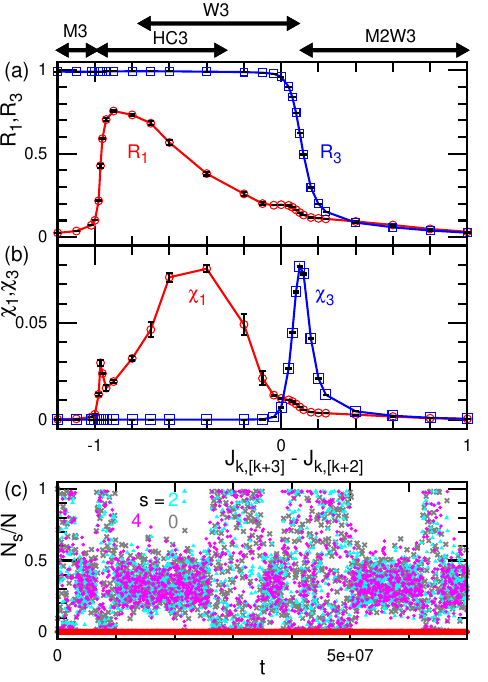}
\caption{
Dependence on $J_{k,[k+3]}-J_{k,[k+2]}$ in the six-state Potts model at $J_{k,[k+2]}+J_{k,[k+3]}=1$, $h=1$,
 and $L=128$.
(a) Order parameters $R_1$ and $R_3$.
The bidirectional arrows at the top represent the ranges of dynamic modes, including the coexistence states.
(b) Susceptibilities $\chi_1$ and $\chi_3$.
The conditions of $J_{k,[k+3]}-J_{k,[k+2]}=-1$, $0$, and $1$
correspond to the points in the phase diagram of Figs.~\ref{fig:6y}, \ref{fig:6d}, and \ref{fig:6t} (Pot6b, Pot6a, and Pot6c), respectively.
(c) Time development of number densities of states at $J_{k,[k+3]}-J_{k,[k+2]}=-0.4$.
The HC3 and W3 modes temporally coexist.
}
\label{fig:m1q}
\end{figure}

\begin{figure}[tbh]
\includegraphics[]{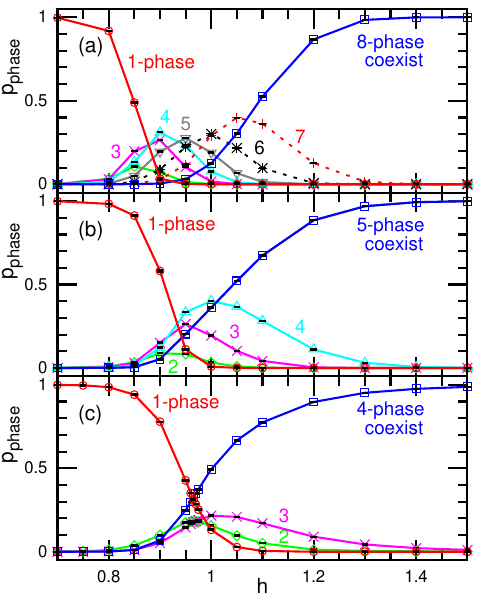}
\caption{
Probabilities $p_{\mathrm{phase}}$ of one-phase dominant and multi-phase coexistence states as functions of $h$
 at $J_{k,k'}=2\delta_{k,k'}$ and $L=128$ for the $q$-state Potts models with
 (a) $q=8$,   (b) $q=5$, and (c) $q=4$.  
The plot in (c) is reproduced from Ref.~\onlinecite{nogu25}. Licensed under CC BY.
}
\label{fig:md01s48}
\end{figure}

\subsubsection{Modes for $J_{k,[k+2]}+J_{k,[k+3]}=1$}\label{sec:6q}

Several modes are shown in the three phase diagrams (Figs.~\ref{fig:6d}(e), \ref{fig:6y}(c), and \ref{fig:6t}(c)).
At the final part of this analysis of the six-state Potts model, the intermediate conditions between these phase diagrams are considered. 
In detail, $J_{k,[k+3]}-J_{k,[k+2]}$ is varied while $J_{k,[k+2]}+J_{k,[k+3]}=1$ is maintained, as shown in Fig.~\ref{fig:m1q}.
As a result, the M3 mode obtained at $J_{k,[k+3]}-J_{k,[k+2]}=-1$ ($J_{k,[k+2]}=1$ and $J_{k,[k+3]}=0$) sequentially changes to
the HC3 mode at $J_{k,[k+3]}-J_{k,[k+2]}=-0.8$ ($J_{k,[k+2]}=0.9$ and $J_{k,[k+3]}=0.1$), 
the W3 mode at $J_{k,[k+3]}-J_{k,[k+2]}=0$ ($J_{k,[k+2]}=0.5$ and $J_{k,[k+3]}=0.5$), and
the M2W3 mode at $J_{k,[k+3]}-J_{k,[k+2]}=1$ ($J_{k,[k+2]}=0$ and $J_{k,[k+3]}=1$). 
The transitions between M3 and HC3 and between HC3 and W3 are characterized by the order parameter $R_1$.
The M3--HC3 transition is second-order, as described in Sec.~\ref{sec:6y}.
At the HC3--W3 transition, the two modes temporally coexist (Fig.~\ref{fig:m1q}(c)), 
and the susceptibility $\chi_1$ has a broad peak (Fig.~\ref{fig:m1q}(b)),
like in the first-order transition at equilibrium.
The W3--M2W3 transition is likely second-order and is characterized by the order parameter $R_3$.
The latter two transitions (HC3--W3 and W3--M2W3) have not seen the aforementioned phase diagrams.
The transitions between other modes (e.g., M3--W3) may be revealed when the full parameter space is explored.

\subsection{Five-State Potts Model}\label{sec:5}

When the interactions of the standard Potts model with $q$ states are used ($J_{k,k'}=2\delta_{k,k'}$),
HC$q$ and W$q$ modes are obtained at low and high $h$, respectively, for $3\leq q\leq 8$ (see Fig.~\ref{fig:md01s48},
the data at $q=7$ are not shown).
When $q>4$,
the WI mode emerges for an intermediate value of $h$,
whereas a direct transition of two modes occurs for $q=3$ and $4$~\cite{nogu24a,nogu25}.
Therefore, the dynamic behaviors of the five-state Potts model with $J_{k,[k+2]}=0$ 
are similar to those of the six-state model with $J_{k,[k+2]}=J_{k,[k+3]}=0$.

\begin{figure}[t]
\includegraphics[]{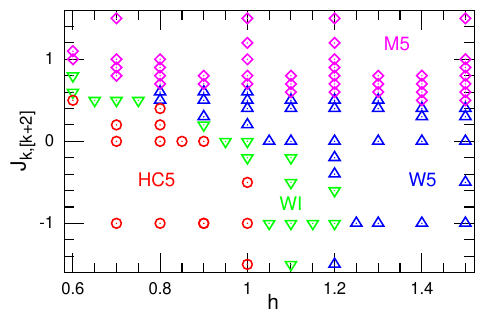}
\caption{
Dynamic phase diagram of the five-state Potts model at $L=128$.
The red circles, magenta diamonds, blue up-pointing triangles, and
green down-pointing triangles represent
HC5, M5, W5, and WI, respectively.
}
\label{fig:5d}
\end{figure}

\begin{figure}[tbh]
\includegraphics[]{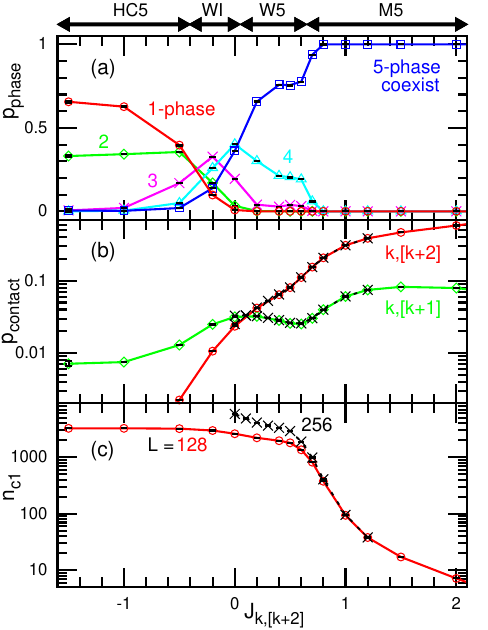}
\caption{
Dependence on $J_{k,[k+2]}$ in the five-state Potts model at $h=1$.
(a) Probabilities $p_{\mathrm{phase}}$ of one-phase and multi-phase coexistence states.
The bidirectional arrows at the top represent the ranges of dynamic modes.
(b) Contact probabilities $p_{\mathrm{contact}}$ for $k$--$[k+1]$ and $k$--$[k+2]$ pairs.
(c) Mean cluster size $n_{\mathrm{c1}}$.
The solid lines represent the data at $L=128$.
In (b) and (c),
the dashed lines with crosses represent the data at $L=256$.
 }
\label{fig:n5m1d}
\end{figure}

The dynamic phase diagram of the five-state Potts model is shown in Fig.~\ref{fig:5d}. 
Since $5$ is a prime number, no factorized-symmetry modes exist.
At $J_{k,[k+2]} \simeq J_{k,k}=2$, a five-state mixed phase (M5) exists, as for the six-state Potts model.
To characterize the transition between the M5 and W5 modes,
we calculated  $p_{\mathrm{phase}}$, $p_{\mathrm{contact}}$, and the mean cluster size $n_{\mathrm{c1}}$, as shown in Fig.~\ref{fig:n5m1d};
At $J_{k,[k+2]}\simeq 0.6$, a stepwise change is observed for $p_{\mathrm{phase}}$ of five states.
Additionally, a minimum is observed for $p_{\mathrm{contact}}$ of $k$--$[k+1]$ pairs and
a sharp decrease in $n_{\mathrm{c1}}$ is initiated.
For $J_{k,[k+2]}\lesssim 0.6$, $n_{\mathrm{c1}}$ is dependent on the system size; however, no size dependence is detected for $J_{k,[k+2]}\gtrsim 0.6$.
In contrast, no size dependence is obtained for $p_{\mathrm{contact}}$  in either range (compare the curves for $L=128$ and $256$ in Fig.~\ref{fig:n5m1d}(b) and (c)).
Thus, we consider that $J_{k,[k+2]} \simeq 0.6$ is the transition point for $h=1$. 
The minimum of $p_{\mathrm{contact}}$ is used to determine the transition points for other values of $h$,
since it is clearer than the other two features.

\subsection{Four-State Potts Model}\label{sec:4}

We first describe phase behavior of the four-state Potts model at thermal equilibrium.
The four-state Potts model corresponds to the standard (two-color) Ashkin-Teller model,
in which two Ising spins, $\sigma^{\mathrm{A}}=\pm 1$ and $\sigma^{\mathrm{B}}=\pm 1$, exist at each lattice site.
The four types of pair states, i.e., $(\sigma^{\mathrm{A}},\sigma^{\mathrm{B}}) =(1,1), (-1,1), (-1,-1)$, and $(1,-1)$,  correspond to
the four states $s=0$, $1$, $2$, and $3$ of the Potts model, respectively.
The phase behavior at thermal equilibrium has been investigated in detail~\cite{ditz80,plas86,bena88}.
Under the conditions $J_{k,k}>J_{k,[k+1]}$ and $J_{k,k}>J_{k,[k+2]}$, 
three phases exist:
a disordered (paramagnetic) phase with $\langle \sigma^{\mathrm{A}} \rangle=\langle \sigma^{\mathrm{B}} \rangle=\langle \sigma^{\mathrm{A}}\sigma^{\mathrm{B}} \rangle=0$,
a Baxter (ferromagnetic) ordered phase with $\langle \sigma^{\mathrm{A}} \rangle \ne 0$, $\langle \sigma^{\mathrm{B}} \rangle \ne 0$, and $\langle \sigma^{\mathrm{A}}\sigma^{\mathrm{B}} \rangle\ne 0$,
and a partially ordered phase with $\langle \sigma^{\mathrm{A}} \rangle=\langle \sigma^{\mathrm{B}} \rangle=0$ and $\langle \sigma^{\mathrm{A}}\sigma^{\mathrm{B}} \rangle\ne 0$.
In the partially ordered phase, two diagonal states ($s=0$ and $2$ or $s=1$ and $3$) are spatially mixed in the view of the Potts model,
which corresponds to the M2 phase for the factorized number of states ($4=2\times 2$) in the present study.

\begin{figure}[tbh]
\includegraphics[]{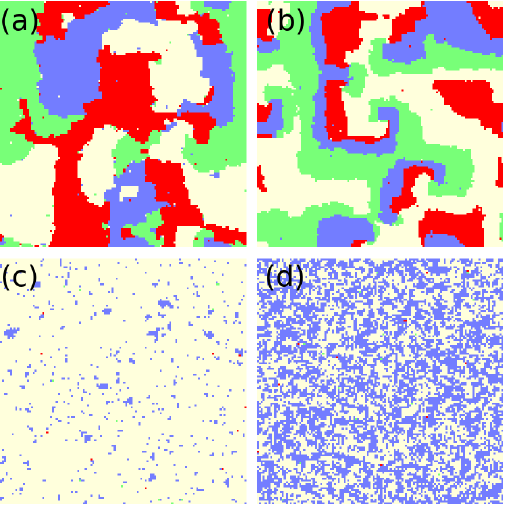}
\includegraphics[]{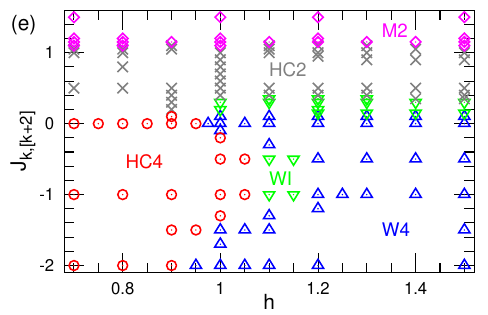}
\caption{
Four-state Potts model at $L=128$.
(a--d) Snapshots.
The light yellow, green, blue, and red sites (light to dark in grayscale) 
represent $s=0$, $1$, $2$, and $3$, respectively.
(a) Nonspiral waves (W4) at  $h=1.1$ and $J_{k,[k+2]}=0$.
(b) Spiral waves (W4) at  $h=1$ and $J_{k,[k+2]}=-2$.
(c) Homogeneous cycling of two diagonal states (HC2) at  $h=1$ and $J_{k,[k+2]}=1$.
(d) Mixing of two diagonal states (M2) at  $h=1$ and $J_{k,[k+2]}=1.5$.
(e) Dynamic phase diagram.
The red circles, gray crosses, magenta diamonds, blue up-pointing triangles, and
green down-pointing triangles represent
HC4, HC2, M2, W4, and WI, respectively.
}
\label{fig:4d}
\end{figure}

\begin{figure}[tbh]
\includegraphics[]{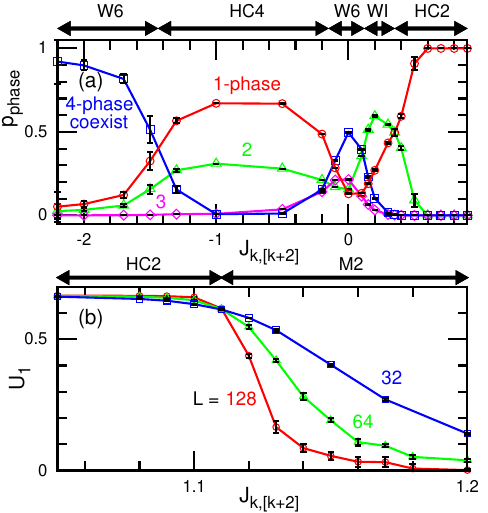}
\caption{
Dependence on $J_{k,[k+2]}$ in the four-state Potts model at $h=1$.
(a) Probabilities $p_{\mathrm{phase}}$ of one-phase and multi-phase coexistence states.
(b) Binder cumulant $U_1$ at $L=32$, $64$, and $128$.
The bidirectional arrows at the top represent the ranges of dynamic modes.
}
\label{fig:n4m1d}
\end{figure}

Here, under active flipping of $h>0$, the M2 phase persists when $J_{k,[k+2]} \gtrsim 1$ (see Fig.~\ref{fig:4d}(d) and (e)).
For intermediate values of $J_{k,[k+2]}$, homogeneous cycling (switching) of two diagonal phases (HC2) emerges (see Fig.~\ref{fig:4d}(c) and (e)).
In the $k$-state dominant phase, the nuclei of $[k+1]$ state are swallowed up by the $[k+2]$ domains.
In the M2 mode, no switching between even- and odd-numbered states occur in the considered simulation periods.
These behaviors are similar to those of the M3 and HC3 modes of the six-state models.
The transition between the HC2 and M2 modes is continuous, and
the transition point is detected based on $U_1$ (see Fig.~\ref{fig:n4m1d}(b)).
Since $R_1 = 0.57\pm 0.01$ at the transition point for $h=1$,
we consider the system is taken as being in the M2 mode when $R_1 < 0.57$.

For $J_{k,[k+2]} \le 0$, 
the HC4 and W4 modes manifest under low and high $h$, respectively.
The WI mode appears only at $J_{k,[k+2]} \simeq -1$, and
the direct HC4--W4 transition occurs through the temporal coexistence at $J_{k,[k+2]} \simeq 0$ and $J_{k,[k+2]} \simeq -2$ (see Figs.~\ref{fig:4d}(e) and \ref{fig:n4m1d}(a)).
The waves form spiral shapes at $J_{k,[k+2]} \simeq -2$,
whereas waves move separately without centers  at $J_{k,[k+2]} \simeq 0$ (see Fig.~\ref{fig:4d}(a) and (b)).
The HC4 and HC2 modes are distinguished by the lifetime ratio and the state-density distribution,
similar to the HC6 and HC3 modes in the six-state models.

\begin{figure}[tbh]
\includegraphics[]{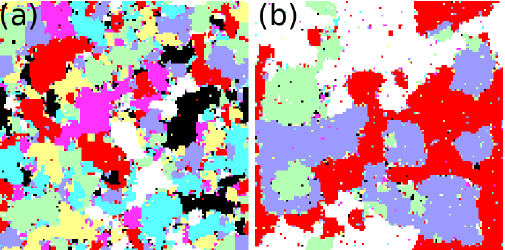}
\includegraphics[]{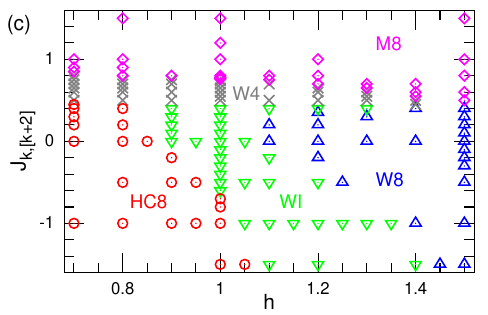}
\caption{
Eight-state Potts model at  $J_{k,[k+3]}=J_{k,[k+4]}=J_{k,[k+2]}$ and $L=128$.
(a) Snapshot of waves of eight states (W8) at  $h=1.3$ and $J_{k,[k+2]}=0$.
(b) Snapshot of waves of four states (W4) at  $h=1$ and $J_{k,[k+2]}=0.6$.
The white, yellow, light green, cyan, blue, magenta, red, and black sites (light to dark in grayscale) 
represent $s=0$, $1$, $2$, $3$, $4$, $5$, $6$, and $7$, respectively.
(c) Dynamic phase diagram.
The red circles, gray crosses, magenta diamonds, blue up-pointing triangles, and
green down-pointing triangles represent
HC8, W4, M8, W8, and WI, respectively.
}
\label{fig:8d}
\end{figure}

\subsection{Eight-State Potts Model}\label{sec:8}

The eight-state Potts model corresponds to the three-color Ashkin-Teller model~\cite{gres81}.
We obtained five modes at  $J_{k,[k+2]}=J_{k,[k+3]}=J_{k,[k+4]}$: HC8, WI, W8, W4, and M8 (see Fig.~\ref{fig:8d}(c)).
The factorized number of states ($8=4\times 2$) are stabilized in the wave mode (W4), in which four domain types ($s=0, 2, 4$, and $6$ or $s=1, 3, 5$, and $7$) spatially coexist.
In the W4 (W8) modes, the boundaries of the $k$ and $[k+2]$ ($[k+1]$) domains move in the direction from the $[k+2]$ ($[k+1]$) domain to the $k$ domains (see Fig.~\ref{fig:8d}(a) and (b) and Movies S5 and S6).

\begin{figure}[tbh]
\includegraphics[]{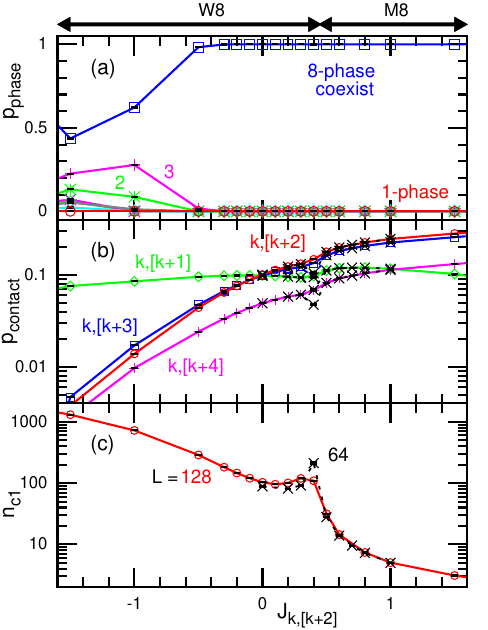}
\caption{
Dependence on $J_{k,[k+2]}$ in the eight-state Potts model at  $J_{k,[k+3]}=J_{k,[k+4]}=J_{k,[k+2]}$ and $h=1.5$.
(a) Probabilities $p_{\mathrm{phase}}$ of one-phase and multi-phase coexistence states.
The bidirectional arrows at the top represent the ranges of dynamic modes.
(b) Contact probabilities $p_{\mathrm{contact}}$ for $k$--$[k+1]$, $k$--$[k+2]$, $k$--$[k+3]$, and $k$--$[k+4]$ pairs.
(c) Mean cluster size $n_{\mathrm{c1}}$.
The solid lines represent the data at $L=128$.
In (b) and (c),
the dashed lines with crosses represent the data at $L=64$.
}
\label{fig:n8m15d}
\end{figure}

The W8 and M8 modes can be distinguished by the contact probabilities $p_{\mathrm{contact}}$ or the mean cluster size $n_{\mathrm{c1}}$ (see Fig.~\ref{fig:n8m15d}(b) and (c)).
Because $n_{\mathrm{c1}}$ exhibits a clear peak (or shoulder) at $J_{k,[k+2]}=0.4$,
we conclude that the transition occurs at this peak point.

In general, the factorized symmetry can emerge when the domain contact of factorized number of states is more stable than the other contacts. For $q=2n$, the contact of even or odd number of states ($s= 2n'$ or $2n'+1$) can be stabilized by increasing $J_{k,[k+2]}$.
Similarly,  $s=3n'$, $3n'+1$, or $3n'+2$ by increasing $J_{k,[k+3]}$ for $q=3n$.
Therefore, the waves of factorized number of states are likely formed also for $q>8$.

\section{Relation with Other Models}\label{sec:other}

The Avni model~\cite{avni25,avni25a} is an active four-state vector Potts model.
They added a nonreciprocal interaction to the Ashkin-Teller model with two Ising spins ($\sigma^{\mathrm{A}}$and $\sigma^{\mathrm{B}}$).
The energy of spin $\sigma^{\alpha}$ in the $i$-th site is given by
\begin{equation}
\label{eq:hint}
E_i^{\alpha} = - J\sum_{\langle ij\rangle} \sigma_i^{\alpha}\sigma_j^{\alpha} - K\epsilon_{\alpha\beta}\sigma_i^{\alpha}\sigma_i^{\beta},
\end{equation}
where $J,K>0$ and $\epsilon_{\alpha\beta}$ is the Levi-Civita symbol. 
The energy of the first term decreases when the spins of the same type in the neighboring sites are in ferromagnetic order.
However, the second term gives  a nonreciprocal intra-site interaction,
such that spin A is preferentially antiferromagnetic to spin B at the same site ($\sigma_i^{A}\sigma_i^{B}<0$), 
whereas spin B is preferentially ferromagnetic ($\sigma_i^{A}\sigma_i^{B}>0$).
Thus, the following cyclic change is induced: $(\sigma^{\mathrm{A}},\sigma^{\mathrm{B}})= (1,1)\to (-1,1) \to  (-1,-1)  \to  (1,-1) \to  (1,1)$.
When the Avni model is mapped to the present active Potts model,
$h =2K$, $J_{k,k} =2J$, $J_{k,[k+1]} =0$, and $J_{k,[k+2]} =-2J$ are obtained.
The contact energy is expressed as $J_{k,[k+j]} =2J\cos(\pi j/2)$, such that these interactions correspond to those of the four-state vector Potts model (also called the clock model)~\cite{wu82}.
In the phase diagram of Fig.~\ref{fig:4d}(e), the data obtained at $J_{k,[k+2]} =-2$ correspond to those of the Avni model at $J=1$.
At low and high $h$, the HC4 and W4 modes occur, respectively, and the two modes temporally coexist around the transition point in the long-term dynamics.

Lattice Lotka--Volterra models, based on predator--prey interactions,
 have been widely used to study spatiotemporal patterns~\cite{szol14,szab02,reic07,szcz13,kels15,dobr18,szab04,szab08,roma12,rulq14,baze19,zhon22,yang23,szol23}.
Cyclic changes in species (states) are considered similar to those of the present models.
The Lotka--Volterra models of four or more species can exhibit short-term dynamics similar to those given by the present model,
such as the long-lived spatial coexistence of two phases and domains of mixed species~\cite{szol14,dobr18,szab04,szab08,roma12,rulq14,baze19,zhon22,yang23,szol23}.
However, the long-term dynamics completely differ.
In the long-term limit, the Lotka--Volterra models exhibit only a uniform phase occupied by a single species through an absorbing transition.
Extinct species never reappear, because species multiply through self-reproduction.
Technically, this is proceeded by pair updates, 
such as $(s_i,s_j)=(0,1) \to (1,1)$, in which the $s_i$ species becomes  the $s_j$ species, indicating predation and self-reproduction.
These updates involve no backward processes; therefore, a detailed balance cannot be obtained.
We compared the single update with this pair update using a three-state Potts model 
and demonstrated that the pair update generates fewer spirals at high flipping energy, but all sites are occupied by a single state in the long-term limit 
under all conditions (including $h=J_{ss'}=0$)~\cite{nogu24a}.
In the case that flips between  neighboring states are only allowed for $q \ge 4$, 
a pattern becomes frozen when the contact of neighboring states disappears.
However, the absorbing transition to a uniform state occurs when flips between non-neighboring states with $h_{kk'}=0$ are added ($k'=[k+2] or [k+3]$ for $q=6$).
Therefore, an update scheme including backward processes is crucial to obtain long-term spatiotemporal dynamics.

\section{Summary}\label{sec:sum}

We studied the nonequilibrium dynamics of the $q$-state active Potts models with $q=4$, $5$, $6$, and $8$ under cyclically symmetric conditions in the long-term limit.
Three mode types and their combinations were identified: mixed phases, wave propagation modes, and cycling of homogeneous phases.
In addition to these modes comprising $q$ states, 
modes comprising a factorized number of states
 emerge for $q=4$, $6$, and $8$: a mixed phase and homogeneous switching of two diagonal states for $q=4$, a wave mode of four states for $q=8$,
and several modes for $q=6$.
In the six-state model, two types of spiral waves with factorized symmetry were found:
spiral waves of three states (three odd numbers, $s=1, 3$, and $5$, or three even numbers, $s=0, 2$, and $4$), 
with the other three states present as small clusters at the domain boundaries;
and spiral waves of three types of mixed domains, in which two diagonal states ($s=k$ and $[k+3]$) are mixed within a domain.
In the former and latter spirals,  one flip cycle in the spiral domains requires one cycle and half cycle of states, respectively.
Moreover, homogeneous cycling modes of the three odd- or even-numbered states and three mixed phases were found.
Therefore, the factorization of symmetry can generate rich dynamic modes under nonequilibrium conditions,
including the combinations of static (mixing) and dynamic (wave) modes.

We calculated the scaling exponents of the second-order transitions between the W3 and M6 modes and between the HC3 and M3 modes.
The exponents of the W3--M6 transition are modified from the equilibrium values,
whereas no change is detected for the HC3--M3 transition.
The transitions to wave modes may generally exhibit greater changes in the exponents than those between static modes.
In contrast,  first-order transitions have  not been reported yet at $N\to\infty$ in the present models.
Temporal coexistences of the HC and wave modes occur, but
the transition points shift to $h=0$ with increasing system size.
Since the forward and backward transitions typically take different pathways, they can exhibit different $N$ dependencies.
However, first-order transitions can be obtained in asymmetric conditions~\cite{nogu25x}.

In this study, we have kept the cyclic symmetry and fixed the temperature and contact energies between the same states and between neighboring states.
In our previous studies, we reported non-cyclic homogeneous phases and the coexistence of two diagonal phases under asymmetric conditions
 (for three- and four-state Potts models)~\cite{nogu24b,nogu25}.
For $q \ge 5$, the imposition of asymmetric conditions will likely modify the dynamic modes in a similar manner; however, symmetry factorization may additionally induce 
different types of dynamics.
Even restricting to the symmetric conditions, we have not explored the entire parameter space.
Under equilibrium, several antiferromagnetic  phases have been reported for the Ashkin-Teller models~\cite{ditz80,gres81,plas86,bena88}.
Thus, antiferromagnetic and other mode types likely exist in the remaining parameter space in active Potts models.
Further, the modes and transitions can be changed  in three (or higher)-dimensional space.
Overall, the active Potts models provide a simple but rich system for exploring various types of nonequilibrium dynamics under thermal fluctuations.

\begin{acknowledgments}
We thank Jun Takahashi for fruitful discussion.
The simulations were partially carried out at ISSP Supercomputer Center, University of Tokyo (ISSPkyodo-SC-2025-Ca-0049).
This work was supported by JSPS KAKENHI Grant Number JP24K06973. 
\end{acknowledgments}


%

\end{document}